\documentclass[english]{paper}
\usepackage[T1]{fontenc}
\usepackage[latin9]{inputenc}
\usepackage{float}
\usepackage{amssymb}
\usepackage{graphicx}
\usepackage{setspace}
\usepackage{esint}
\makeatletter
\newcommand{\lyxaddress}[1]{
\par {\raggedright #1
\vspace{1.4em}
\noindent\par}
}

\makeatother

\usepackage{babel}
\begin{document}

\title{Quantum Financial Economics - Risk and Returns}

\author{Carlos Pedro Gonçalves}

\institution{Instituto Superior de Ciências Sociais e Políticas (ISCSP), Technical
University of Lisbon}

\maketitle

\lyxaddress{E-mail: cgoncalves@iscsp.utl.pt}
\begin{abstract}
Financial volatility risk and its relation to a business cycle-related
intrinsic time is addressed through a multiple round evolutionary
quantum game equilibrium leading to turbulence and multifractal signatures
in the financial returns and in the risk dynamics. The model is simulated
and the results are compared with actual financial volatility data.\end{abstract}
\begin{keywords}
Quantum Financial Economics, Business Cycle Dynamics, Intrinsic Time,
Quantum Chaotic Volatility
\end{keywords}

\section{Introduction}

Ever since Mandelbrot identified the presence of multifractal turbulence
in the markets \cite{key-15,key-16,key-17}, this empirical fact has
become a major research problem within financial economics.

Mandelbrot \cite{key-15,key-16} hypothesized that financial systems'
dynamics has to be addressed in terms of an intrinsic temporal notion,
linked to the economic rhythms and (chaotic) business cycles. Such
intrinsic time would not be measured in clock time, but in terms of
economic rhythms that would rescale volatility with the usual square
root rule that holds for clock-based temporal intervals.

In the present work, we return to such a proposal, providing for a
quantum game theoretical approach to market turbulence with chaotic
intrinsic time leading to multifractal signatures in volatility dynamics.
The approach followed is that of path-dependent quantum adaptive computation
within the framework of quantum game theory, such that a game is divided
in rounds and, for each round, an equilibrium condition is formalized
in terms of a payoff quantum optimization problem, subject to: (1)
a time-independent Schrödinger equation for the round; (2) an update
rule for the Hamiltonian, depending on some evolutionary parameter(s)%
\footnote{The time-independent Schrödinger equation can be addressed either
as attaching an eigenstate to the whole round, for the game's result,
or as assigning it to the round's end, and the change in the parameters
leads to a change in the time-independent equation for the round,
given the previous round. In \cite{key-9}, such approach with discrete
game rounds was also considered, with unitary evolution between each
two rounds. In the present case, instead of a unitary evolution operator,
we have a quantum optimization problem per round, leading to a quantum
strategy formulation. For the game proposed in \cite{key-9} the two
approaches are, actually, equivalent, since they form part of the
underlying approach to the path-dependent quantum computation approach
to quantum games.%
}.

The Hamiltonian for the time-independent Schrödinger equation constitutes
a quantum evolutionary expansion of the classical harmonic oscillator
approaches to the business cycle dynamics in economics \cite{key-11,key-8,key-24,key-19,key-7},
thus, leading to a quantum business cycle approach to business financial
valuation by a financial market, such that a clock time independent
quantum state is associated with each round, where business-cycle
related financial intrinsic time results from the quantum game itself,
without any stochastic temporal subordination over clock time \cite{key-15,key-16}.

The model, therefore, incorporates the business cycle by adapting
the standard economic tradition, in business cycle dynamics modelling,
in particular, extending to the quantum setting a standard harmonic
oscillator model of the business cycle, within the context of an adaptive
quantum business optimization problem with nonlinear evolutionary
conditions.

The present model's main objective is to translate into a quantum
repeated game setting what consitutes Mandelbrot's hypothesis of economic
complexity with financial market efficiency, such that turbulence
and multifractal signatures do not come from anomalous trading behavior
or speculative trading systems, but from the business cycle nonlinear
dynamics, that is, from underlying economic evolutionary dynamics
affecting a company's fundamental value in a value efficient financial
market, such that trading accurately reflects that underlying economic
business cycle dynamics in the markets' turbulent volatility dynamics.

Our present goal is, therefore, to contribute to the discussion within
financial economics, regarding the issue of business cycle-related
trading time, raised by Mandelbrot as a criticism to the empirical
validity of the geometric Brownian motion and geometric random walk
models of price dynamics, used in financial economics as mathematical
tools for solving pricing problems.

In \emph{section 2.}, we provide for a brief review of financial economics
and quantum financial economics, laying down the background to the
present work in its connection with other works and with the general
field of financial economics. In \emph{section 3.}, we introduce the
model and address the main findings from the model's simulation. In
\emph{section 4.}, we address the implications of the model for financial
economics.

\section{Financial Economics}

Financial economics is a branch of economic theory which deals with
a combination of mathematical finance tools with economic theories,
applying these to the context of financial problems dealing with time,
uncertainty and resource management (in particular allocation and
deployment of economic resources), as stressed by Merton \cite{key-18}.

Earlier models in quantum financial economics have worked with quantum
Hamiltonian proposals, including harmonic oscillator potentials and
applications of quantum theory to option pricing, most notably one
may quote: Segal and Segal \cite{key-30}, which constitutes one of
the earliest works in quantum option pricing theory, as well as Baaquie's
quantum path-integral approach to option pricing with stochastic volatility
\cite{key-1,key-2,key-3,key-4}, who largely divulged the quantum
formalism to the financial economics community, by applying it to
problems that are specific of that research area, this is explored
in depth in the work \cite{key-3}, with the presentation and explanation
of path-integral examples taken from quantum mechanics, including
the quantum harmonic oscillator, and with the application to several
examples from financial economics. Simultaneously to Baaquie, still
within the specific area of quantum applications to financial economics,
one may also quote the work of Schaden \cite{key-27,key-28,key-29},
to name but a few of the early works in the field of quantum financial
economics.

Along with quantum financial economics one may also refer the parallel
and related research field of quantum financial game theory, specifically
the work of Piotrowski and Sladkowski \cite{key-20,key-21,key-22}
who applied quantum game theory to financial theory, and Gonçalves
and Gonçalves \cite{key-9}, who proposed and tested empirically a
model of a quantum artificial financial market.

Regarding the quantum harmonic oscillator approach, as a specific
formulation within quantum financial game theory, one may refer both
Ilinsky \cite{key-12}, who addressed a financial interpretation of
the quantum harmonic oscillator, as well as the above quoted work
of Piotrowski and Sladkowski \cite{key-20,key-21,key-22}. Regarding
Piotrowski and Sladkowski's work, the main difference, in regards
to the current approach is that the quantum harmonic oscillator proposed
in \cite{key-20,key-21,key-22} resulted from a financial operator
related to speculators' buying and selling strategies and risk profiles,
while, in our case, it is a quantum extension based upon a tradition
of non-quantum economics business cycle modelling, such that the quantum
Hamiltonian, in our case, is a company's strategic operator related
to that company's business value dynamics and the quantum optimization
problem is linked to the company's economic risk management process,
which is consistent with the fact that we are dealing with business
economic risk linked to the business cycle that is reflected in the
stock market price by value investors, thus, leading to a value efficient
financial market, as per Mandelbrot's hypothesis of a financial system
which is efficient in intrinsic time \cite{key-15,key-16,key-17}.

Therefore, while the quantum Hamiltonian operator proposed in \cite{key-20,key-21,key-22}
constitutes a risk inclination operator and the (financial) mass term
corresponds to a financial risk asymmetry in buying and selling strategies,
in the model addressed here, the mass term corresponds to a round-specific
measure of business economic evolutionary pressure divided by business
cycle frequency, therefore, it has an economic business-related interpretation,
similarly the harmonic oscillator oscillation frequency corresponds
to the business cycle oscillation frequency.

All the variables in the optimization problem, that are worked with
in the present model, stem from an expansion of basic business cycle
evolutionary economic dynamics to the quantum setting. Therefore,
the roots of the present model lie in the classical mechanics-based
business cycle economics, addressed from an evolutionary perspective
with contra-cyclical adaptive dynamics, resulting from Püu's review
and from proposals regarding chaotic dynamical models for the business
cycle \cite{key-24,key-10,key-13}.

Having circumscribed the approach within the appropriate literature,
we are now ready to introduce the model.

\section{A Quantum Financial Game and Quantum Financial Economics}

Let $S_{t}$ be the financial market price of a company's shares,
transactioned synchronously by traders, in discrete rounds, at the
end of each round, and let $r_{t}$ be a rate of return, such that:
\begin{equation}
S_{t}=S_{t-\triangle t}e^{r_{t}}
\end{equation}
with:
\begin{equation}
r_{t}=\ln\left(\frac{S_{t}}{S_{t-\triangle t}}\right)=\mu\triangle t+\sigma x_{t}
\end{equation}
where $\mu$ is a fixed average return, $\triangle t$ is the duration
of a game's round, $\sigma$ is a fixed volatility component.

The subscript $t$ labels the round in accordance with its final transaction
time, $t=\triangle t,2\triangle t,3\triangle t,...$, as is the usual
framework in game theory for a repeated game, where each round corresponds
to an iteration of the game with the same game conditions (\emph{fixed
repeated game}) or with evolving conditions (\emph{evolutionary repeated
game}).

Considering a financial market composed by value investors, it is
assumed that market participants accurately evaluate the company's
fundamental value such that $r_{t}$ is a fair return on the company's
shares. In this way, the variable $x_{t}=\frac{r_{t}-\mu\triangle t}{\sigma}$
represents a volatility adjusted component representing, in a market
dominated by value investors, a company's \emph{value fitness}, which
is related to the company's business growth prospects.

If we were dealing with classical economic business cycle dynamics,
one might consider the dynamics for $x_{t}$ to be driven by the harmonic
oscillator potential $V(x)=-\frac{x^{2}}{b}$, where the parameter
$b$ is the business evolutionary pressure, which includes the ability
of the company to quickly adapt to adverse economic conditions, as
well as increased business growth restrictions, such that: the higher
the value of $b$ is, the more competitive is the business environment.
Unlike in physics, within the economic setting, the parameter $b$
is considered dimensionless.

In the economic potential $V(x)=-\frac{x^{2}}{b}$ we have that: $x<0$
signals negative factors and possibly a downward period in the business
cycle dynamics, while $x>0$ signals positive factors and possibly
business grown, thus, in the harmonic oscillator potential, it is
assumed that negative $x$ is dampened by actions on the part of the
company towards the recovery, while positive $x$ may be dampened
by business growth restrictions, which includes competition with other
companies.

The existence of such contra-cyclical dynamics, affecting both positive
and negative business cycle processes is formalized by the harmonic
oscillator potential. Such potentials have been widely used within
the economic modelling of business cycles, appearing in the dynamics
of the multiplier and accelerator as well as in the Frischian tradition
to the business cycle modelling \cite{key-11,key-8,key-24,key-19,key-7}.
In the present case, the evolutionary interpretation comes from a
combination of Püu's work \cite{key-24} on the business cycle dynamics
with related chaotic evolutionary coupled map lattice proposals \cite{key-10}.

Within a quantum game setting, we consider a repeated business game.
The general form of a quantum repeated game, addressed here, is such
that the game is divided in rounds, with a fixed Hilbert space, and
at each round an optimization problem is assigned with eigenvalue
restrictions holding for the round. Thus, given an appropriate payoff
operator, the game is completely defined by a sequence of optimization
problems leading to a round-specific sequence of game equilibrium
wave functions which solve the sequence of optimization problems.

In the present case, we introduce the \emph{fitness} operator $\hat{x}$,
satisfying the eigenvalue equation:
\begin{equation}
\hat{x}\left|x\right\rangle =x\left|x\right\rangle 
\end{equation}
Since $x$ ranges in $\mathbb{R}$, it follows that the spectrum is
continuous and the kets $\left|x\right\rangle $ and the bras $\left\langle x\right|$
are not in a Hilbert space, rather, it becomes necessary to work with
a rigged Hilbert space%
\footnote{Well known to the complexity approach to economics' community, in
particular those linked to applications of the Brussels-Austin Schools
and Prigogine's works on complex systems \cite{key-23,key-6}%
} given by the Gelfand triplet $\Phi\subset\mathcal{H}_{G}\subset\Phi^{\times}$,
where \cite{key-14}: $\Phi$ (the space of test functions) is a dense
subspace of the game's Hilbert space $\mathcal{H}_{G}$, $\mathcal{H}_{G}$
arises from the requirement that the wave functions $\psi\in\Phi$,
which correspond to the quantum strategic configurations, be square
normalizable and $\Phi^{\times}$ (the space of distributions) is
the space of antilinear functionals over $\Phi$, such that $\left|x\right\rangle \in\Phi^{\times}$.
Similarly, to address the bras, we have to work with the triplet $\Phi\subset\mathcal{H}_{G}\subset\Phi'$,
where $\Phi'$ is the space of the linear functionals over $\Phi$,
such that \cite{key-14}: $\left\langle x\right|\in\Phi'$. 

This allows for the basic quantum mechanical prescriptions to hold
\cite{key-14}:
\begin{equation}
\left\langle x|x'\right\rangle =\delta(x-x')=\intop_{-\infty}^{+\infty}du\left\langle x|u\right\rangle \left\langle u|x'\right\rangle 
\end{equation}
\begin{equation}
\psi(x)=\left\langle x|\psi\right\rangle =\intop_{-\infty}^{+\infty}du\left\langle x|u\right\rangle \left\langle u|\psi\right\rangle 
\end{equation}

In a repeated quantum game, there is a quantum strategy for each round
of the game, therefore, it becomes natural to index the wave function
by the corresponding round index in order to identify to which round
it belongs, thus, $\psi_{t}(x)$ is the quantum strategic configuration
for the $t$-th round.

Each round's strategic configuration results from a quantum optimization
problem defining a \emph{quantum business game evolutionary equilibrium}.
The optimization is defined in terms of the squared operator $\hat{x}^{2}$,
such that the higher is the round's expected value for $\hat{x}^{2}$,
denoted by $\left\langle \hat{x}^{2}\right\rangle _{\psi_{t}}$, the
higher is the company's economic risk, this means that higher absolute
values for returns correspond to higher financial volatility risk
linked, in this case, to economic volatility risk, while lower returns
in absolute value correspond to smaller volatility risk.

To assume that the company adapts to business economic volatility
risk means that the quantum game equilibrium is defined in terms of
the minimization of risk, that is, the \emph{fitness dispersion }is
kept as low as possible as well as the returns, making risk smaller,
this is achieved by minimizing $\left\langle \hat{x}^{2}\right\rangle _{\psi_{t}}$,
or, alternatively, maximizing its negative $-\left\langle \hat{x}^{2}\right\rangle _{\psi_{t}}$,
since $\max\left\{ -\left\langle \hat{x}^{2}\right\rangle _{\psi_{t}}\right\} =\min\left\{ \left\langle \hat{x}^{2}\right\rangle _{\psi_{t}}\right\} $.
In order to be more straightforward, in regards to an economic interpretation,
we define the optimization game in terms of the company's risk minimization
objective $\min\left\{ \left\langle \hat{x}^{2}\right\rangle _{\psi_{t}}\right\} $,
rather than in terms of the equivalent maximization of a negative
payoff, this leads to the following round-specific economic business-cycle
volatility minimization problem for the company:

\begin{equation}
\begin{array}{cc}
 & \min\left\{ \left\langle \hat{x}^{2}\right\rangle _{\psi_{t}}\right\} \\
s.t. & \hat{H}_{t}\psi_{t}(x)=E\psi_{t}(x)\\
 & \hat{H}_{t}=-\frac{\hbar_{s}^{2}}{2m_{t}}\frac{d^{2}}{dx^{2}}+\frac{b}{2}x^{2}
\end{array}
\end{equation}
The quantum business cycle Hamiltonian operator translates to the
financial economic setting, with a few adaptation in units. Indeed,
energy is, in this case, expressed in units of returns and the shares'
Planck-like constant $\hbar_{s}$ plays a similar role to that of
quantum mechanics' Planck constant, indeed, the quanta of energy for
the quantum harmonic oscillator game's restrictions at round $t$
are:
\begin{equation}
E_{n}(t)=\left(n+\frac{1}{2}\right)\hbar_{s}\omega_{t}
\end{equation}
where $\omega_{t}$ represents the angular frequency of oscillation
of the business cycle for the round $t$, expressed as radians over
clock time%
\footnote{It should be stressed that $\omega_{t}$ is associated with the round
itself, as a part of the game's restrictions and the subscript identifies
the angular frequency as such, and not as a continuous clock time
dependency. One may assume, alternatively, that $\omega_{t}$ is assigned
to the round's end where the decision takes place with a wave function
that results from the optimization problem presented in the text.%
}, and $\hbar_{s}$ is expressed as $\frac{h_{s}}{2\pi}$, where $h_{s}$
is expressed in units of returns over the business cycle oscillation
frequency for the round $t$, such oscillation frequency is, in turn,
obtained from $\nu_{t}=\frac{\omega_{t}}{2\pi}$, thus, being expressed
in terms of the number of business-related oscillation cycles per
clock time.

We also consider a business cycle-related mass-like term which can
be obtained from the relation:

\begin{equation}
\omega_{t}=\left(\frac{b}{m_{t}}\right)^{\frac{1}{2}}
\end{equation}
leading to:
\begin{equation}
m_{t}=\frac{b}{\omega_{t}^{2}}
\end{equation}
thus, since $b$ is dimensionless, the business cycle mass-like term
is expressed in units of inverse squared angular frequency.

Solving, first, for the quantum Hamiltonian restrictions, the feasible
set of quantum strategies is obtained, for the round, as the eigenfunctions
of the quantum harmonic oscillator:
\begin{equation}
\psi_{n,t}(x)=\left(\frac{\alpha_{t}}{\sqrt{\pi}2^{n}n!}\right)^{\frac{1}{2}}e^{-\alpha_{t}^{2}\frac{x^{2}}{2}}H_{n}(\alpha_{t}x)
\end{equation}
\begin{equation}
\alpha_{t}=\sqrt[4]{\frac{m_{t}b}{\hbar_{s}^{2}}}
\end{equation}

The round specific expected risk for each alternative strategy is,
then, given by:

\begin{equation}
\left\langle \hat{x}^{2}\right\rangle _{\psi_{n,t}}=\frac{1}{\alpha_{t}^{2}}\left(n+\frac{1}{2}\right)=\left(n+\frac{1}{2}\right)\frac{\hbar_{s}}{m_{t}\omega_{t}}=\frac{E_{n}(t)}{m_{t}\omega_{t}^{2}}=\frac{E_{n}(t)}{b}
\end{equation}

Minimization of expected risk by the company leads to:
\begin{equation}
\min\left\{ \left\langle \hat{x}^{2}\right\rangle _{\psi_{n,t}}\right\} =\frac{E_{0}(t)}{b}
\end{equation}
Therefore, the quantum game's evolutionary equilibrium strategy is
the eigenfunction for the zero-point energy solution of the quantum
harmonic oscillator: 
\begin{equation}
\psi_{0,t}(x)=\left(\frac{\alpha_{t}}{\sqrt{\pi}}\right)^{\frac{1}{2}}e^{-\alpha_{t}^{2}\frac{x^{2}}{2}}=\left(\frac{1}{\theta_{t}\sqrt{2\pi}}\right)^{\frac{1}{2}}e^{-\frac{x^{2}}{4\theta_{t}^{2}}}
\end{equation}
where $\theta_{t}$ is a business cycle-related volatility parameter
defined as:

\begin{equation}
\theta_{t}=\frac{1}{\sqrt{2}\alpha_{t}}=\sqrt{\min\left\{ \left\langle \hat{x}^{2}\right\rangle _{\psi_{n,t}}\right\} }=\sqrt{\frac{E_{0}(t)}{b}}
\end{equation}
which makes explicit the connection between the quantum game equilibrium
strategy for the round and the risk optimization problem.

Introducing the volatility component $K_{t}$, such that:
\begin{equation}
K_{t}=\frac{E_{0}(t)}{2b}=\frac{\left\langle T\right\rangle _{\psi_{0,t}}}{b}
\end{equation}
that is, $K_{t}$ is equal to the expected value of the quantum harmonic
oscillator's kinetic energy for the round, divided by the evolutionary
pressure constant, thus, $K_{t}$ is called \emph{kinetic volatility
component}. Replacing in $\theta_{t}$, we obtain:

\begin{equation}
\theta_{t}=\sqrt{2K_{t}}
\end{equation}

The final result of this quantum game, for the financial returns,
is the returns' wave function for the game round:
\begin{equation}
\psi_{0,t}(r)=\left(\frac{\alpha_{t}}{\sigma\sqrt{\pi}}\right)^{\frac{1}{2}}e^{-\alpha_{t}^{2}\frac{(r-\mu\triangle t)^{2}}{2\sigma^{2}}}=\left(\frac{1}{\left(\sqrt{2K_{t}}\sigma\right)\sqrt{2\pi}}\right)^{\frac{1}{2}}e^{-\frac{(r-\mu\triangle t)^{2}}{4\left(\sqrt{2K_{t}}\sigma\right)^{2}}}
\end{equation}
In the Gaussian random walk model of financial returns, within neoclassical
financial theory, the following density is assumed \cite{key-16}:
\begin{equation}
dP_{neoclassical}^{\triangle t}=\frac{1}{\left(\sqrt{\triangle t}\sigma\right)\sqrt{2\pi}}\exp\left[\frac{\left(r-\mu\triangle t\right)^{2}}{2\left(\sqrt{\triangle t}\sigma\right)^{2}}\right]dr
\end{equation}
where $\triangle t$ is a discrete time step.

Mandelbrot's proposal is that the business cycle-related intrinsic
time lapse $\tau_{B}(t)$ should be used instead of the clock time
interval of $\triangle t$, the business cycle rhythmic time%
\footnote{Volume, absolute returns or any other relevant such measure have been
used as surrogates for such intrinsic market time, any such notion
that can define a sequence of steps on a devil's staircase can represent
a form of intrinsic time, which does not coincide with clock time
units. Intrinsic time is, thus, assumed to be financial business cycle-related
time which is usually measured in financially relevant units \cite{key-16}.%
} marks a round duration that does not numerically coincide with a
clock time interval, but rather with an intrinsic business cycle time
interval, for the round $t$, denoted by $\tau_{B}(t)$ which affects
the volatility as follows:
\begin{equation}
\tau_{B}(t)=2K_{t}=\frac{E_{0}(t)}{b}=\frac{\hbar_{s}\omega_{t}}{2b}
\end{equation}
thus, the intrinsic time frame is expressed not in clock time but
in units of returns related to the financial energy, as explained
earlier. The intrinsic temporal sequence is, thus, given by $\frac{\hbar\omega_{1\triangle}}{2b},\frac{\hbar(\omega_{1\triangle}+\omega_{2\triangle})}{2b},\frac{\hbar(\omega_{1\triangle}+\omega_{2\triangle}+\omega_{3\triangle})}{2b},...$,
this sequence naturally defines a nondecreasing sequence. When $\omega_{t}$
has a turbulent stochastic behavior, the sequence leads to a devil's
staircase, corresponding to a turbulent intrinsic time related to
the business cycle rhythmic time.

The corresponding Gaussian probability density is, in this case, given
by:
\begin{equation}
dP_{t}=\frac{1}{\left(\sqrt{\tau_{B}(t)}\sigma\right)\sqrt{2\pi}}e^{-\frac{\left(r-\mu\triangle t\right)^{2}}{2\tau_{B}(t)\sigma^{2}}}
\end{equation}
the two temporal notions, that of clock time and that of intrinsic
time, appear in the density. The clock time appears multiplying by
the average returns, since the evidence is favorable that the intrinsic
time is directly related to market volatility rather than to the average
returns%
\footnote{That is, the market seems to evaluate the average returns with a clock
temporal scale, while the volatility scales in intrinsic time, which
is related to the fact that the volatility is linked to transaction
rhythms and to the business cycle risk processing by the markets \cite{key-15,key-16}.%
}.

Turbulence, power-law scaling and multifractal signatures arise, in
the model, from the dynamics of $\tau_{B}(t)$, through the following
power-law map:
\begin{equation}
K_{t}=\left[\frac{\left(1-\varepsilon\right)2u\left(K_{t-\triangle t}^{\frac{1}{1-D}}-1\right)mod\,1+\varepsilon|r_{t-\triangle t}|}{u}+1\right]^{1-D}
\end{equation}
with parameters $0<u<1$ and $0\leq\varepsilon\leq1$. The above map
is conjugate to the coupled shift map:
\begin{equation}
I_{t}=(1-\varepsilon)2I_{t-\triangle t}mod\,1+\varepsilon\left|r_{t-\triangle t}\right|
\end{equation}
through the power law relation defined over the \emph{kinetic volatility
component}%
\footnote{The power law dependency is to be expected, following Mandelbrot's
empirical work, which shows that economic processes seem to lead to
scale invariance in risk dynamics\cite{key-15,key-16,key-17}.%
}:
\begin{equation}
K_{t}=\left(1+\frac{I_{t}}{u}\right)^{1-D}
\end{equation}
From conjugacy with the shift map it follows that, when $\varepsilon=0$,
the following normalized invariant density holds for $K$:
\begin{equation}
\rho(K)=\frac{K^{-\frac{D}{1-D}}}{1-D}
\end{equation}
with normalization achieved from division by $u$. Therefore, $D$
is a scaling power-law parameter related to the volatility statistical
distribution, such that if we write $\alpha:=\frac{D}{1-D}$, and
$\lambda:=\frac{\alpha}{D}$, we obtain the power-law density:
\begin{equation}
\rho(K)=\lambda K^{-\alpha}
\end{equation}
which is consistent with evidence from power-law volatility scaling
in the financial markets \cite{key-15,key-16,key-17}.

The Bernoulli shift map for the dynamics of $I_{t}$ formalizes a
dynamics of business cycle-related expansion and contraction in volatility
conditions with a uniform invariant density%
\footnote{Conceptually, the $I_{t}$ variable can be interpreted as synthesizing
risk factors associated with fundamental value, that is, to fundamental
value risk drivers.%
}. The Bernoulli map is coupled to the previous round's financial returns,
formalizing a feedback from the market itself upon the economic behavior
of the volatility fundamentals $I_{t}$. For a coupling of $\varepsilon\neq0$,
the quantum feedback affects the chaotic map, leading to a situation
in which the previous round's volatility, measured by the absolute
returns, affects the current round's chaotic dynamics.

Taking all of the elements into account, the final quantum game's
structure is given by:

\begin{equation}
\begin{array}{cc}
 & \min\left\{ \left\langle \hat{x}^{2}\right\rangle _{\psi_{t}}\right\} \\
s.t. & \hat{H}_{t}\psi_{t}(x)=E\psi_{t}(x)\\
 & \hat{H}_{t}=-\frac{\hbar_{s}^{2}}{2m_{t}}\frac{d^{2}}{dx^{2}}+\frac{b}{2}x^{2}\\
 & m_{t}=\frac{\hbar_{s}^{2}}{4b\tau_{B}(t)^{2}}\\
 & \tau_{B}(t)=2K_{t}=\frac{\hbar_{s}\omega_{t}}{2b}\\
 & K_{t}=\left[\frac{\left(1-\varepsilon\right)2u\left(K_{t-\triangle t}^{\frac{1}{1-D}}-1\right)mod\,1+\varepsilon|r_{t-\triangle t}|}{u}+1\right]^{1-D}
\end{array}
\end{equation}

In figure 1, is shown the result of a simulation of the quantum financial
game with this structure. The presence of market turbulence can be
seen in the financial returns series, resulting from the Gaussian
density shown in Eq.(21).

\begin{figure}[H]
\begin{centering}
\includegraphics[clip,scale=0.7]{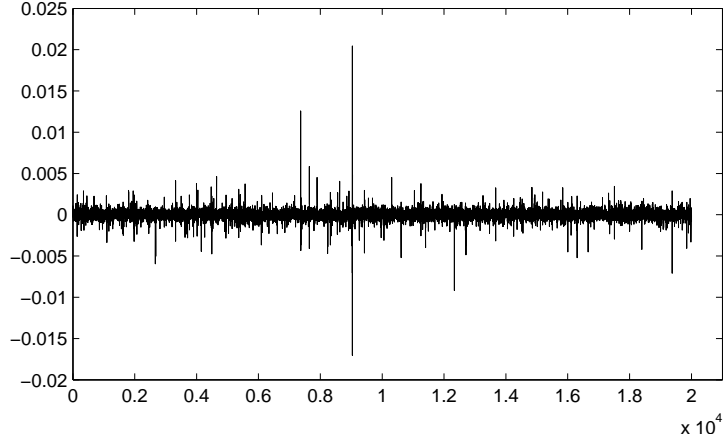}
\par\end{centering}

\caption{Netlogo simulation of the model, with parameters: $\varepsilon=0.001$,
$u=1.0E-5$, $D=1.83$, $\mu=1.0E-6$, $\triangle t=1$ $\sigma=0.02$.
Simulation with 30,000 rounds, the first 10,000 having been removed
for transients.}
\end{figure}

In figure 2, a multifractal large deviation spectrum (estimated with
Fraclab) is presented for the financial returns, showing a peak around
0.5, which is in accordance with Mandelbrot, Fisher and Calvet's hypothesis
of multifractal financial efficiency \cite{key-15,key-16,key-17}.

\begin{figure}[H]
\begin{centering}
\includegraphics[clip,scale=0.5]{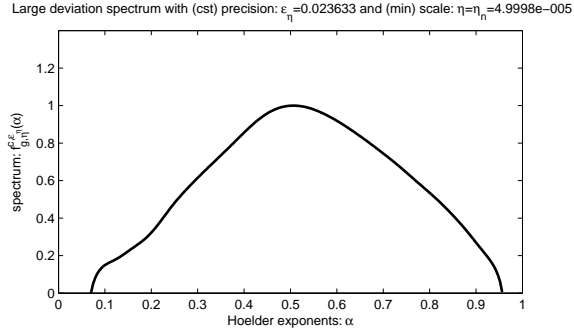}
\par\end{centering}

\caption{Large deviation spectrum for the financial returns, obtained from
a Netlogo simulation of the quantum market game with parameters: $\varepsilon=0.001$,
$u=1.0E-5$, $D=1.83$, $\mu=1.0E-6$, $\triangle t=1$, $\sigma=0.02$.
The spectrum was estimated with 30,000 rounds, the first 10,000 removed
for transients.}
\end{figure}

In figure 3, the multifractal spectra for the dynamics of $K_{t}$
is shown, assuming three different values for the coupling parameter.
The presence of multifractality for $\varepsilon=0$ shows that the
chaotic dynamics is responsible for the emergence of the turbulence
and multifractal scaling, thus, we are dealing with multifractal chaos
with origin in the nonlinear dynamics of business cycle volatility
risk%
\footnote{The presence of chaos in business cycles is a known empirical fact
\cite{key-7}, the current model addresses that chaos in connection
to volatility.%
}.

\begin{figure}[H]
\begin{centering}
\includegraphics[scale=0.5]{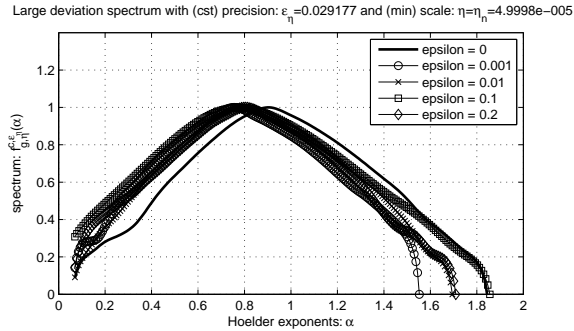}
\par\end{centering}

\caption{Large Deviation Spectra obtained in Fraclab for $K_{t}$, from a Netlogo
simulation of the quantum market game, for different coupling values,
with parameters: $u=1.0E-5$, $D=1.83$, $\mu=1.0E-6$, $\triangle t=1$,
$\sigma=0.02$. The spectra were estimated with 30,000 rounds, the
first 10,000 removed for transients.}
\end{figure}

For $\varepsilon\neq0$, the quantum fluctuations that affect the
dynamics for $K_{t}$ seem to lead to a lower value of the peak of
the multifractal spectrum, indicating a higher irregularity in the
motion. On the other hand, when $\varepsilon=0$, there emerges a
multifractal spectrum with a peak that is closer to 1, showing evidence
of higher persistence and more regular dynamics. For all of the couplings,
however, there is evidence of persistence in the dynamics of $K_{t}$,
which is in accordance with previous findings for the financial markets
and business cycles' empirical data \cite{key-15,key-16,key-17,key-12}.

One can also identify, in the volatility spectra of the simulations,
Hölder exponents larger than 1, which is characteristic of turbulent
processes where there are clusters of irregularity representing short
run high bursts of activity which tend to be smoothed out by laminar
periods in the longer run. This signature is not dominant in the game's
simulations but may take place, which is favorable evidence since
such spectra signatures take place in actual market volatility measures
expressing adaptive expectations regarding volatility fundamentals,
as it is shown in figure 4, for the volatility index {}``VIX'' which
is the volatility index on the S\&P 500.

\begin{figure}[H]
\begin{centering}
\includegraphics[clip,scale=0.5]{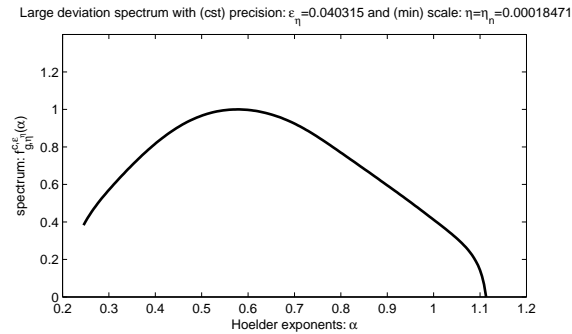}
\par\end{centering}

\caption{Large deviation spectrum estimated in Fraclab for the VIX daily closing
historical values during the period from 02-01-1990 to the period
27-06-2011. The spectrum peaks at a value of Hölder exponent larger
than 0.5 showing evidence of persistence, and there is a region of
scaling with Hölder exponents larger than 1.}
\end{figure}

Even though the $K_{t}$ is not a volatility index, the conceptual
proximity regarding the incorporated expectations allow for some comparison.
The large deviation spectrum of the VIX also shows a lower persistence
which is more consistent with the cases for $\varepsilon\neq0$, a
result to be expected since the financial returns' volatility seem
to be affected by the magnitude of previous returns.

\section{Conclusions}

The present work has combined chaos theory and quantum game theory
to provide for a game theoretic equilibrium foundation to Mandelbrot's
argument of intrinsic time linked to the business cycle as a source
of turbulent dynamics and multifractal signatures in the financial
markets.

If we were to let $2K_{t}=\triangle t$, then, we would trivially
obtain the traditional log-normal random walk model, by letting $2K_{t}=\tau_{B}(t)$
we were able to implement Mandelbrot's proposal of a business cycle-related
intrinsic time, and, thus, to provide for a quantum version of an
evolutionary business cycle approach to financial turbulence.

A significant econometric point of the model regards the volatility
scaling, indeed the multifractal signatures, in this case, result
from the nonlinear dependencies rather than from a prescribed fixed
multifractal measure: we are dealing with power-law conditional heteroscedasticity
responsible for the emergence of multifractal signatures. This allows
one to establish a bridge between Mandelbrot's proposal and conditional
heteroscedasticity models.

From the economic analysis perspective, the game's simulations show
that the interplay between economic chaos and volatility dynamics
may account for the emergence of turbulence and multifractal signatures.
The quantum approach has advantages over the classical stochastic
processes since it provides for theoretical foundations underlying
the probability measures, linking the probability densities to the
underlying game structures and economic dynamics, while sharing the
same advantage of being ammenable to econometric analysis and estimation,
which can prove useful in portfolio management, derivative pricing
and risk management, all areas of application of financial economics.

\end{document}